\documentclass[a4paper,11pt]{article}
\usepackage{comment}

\usepackage[english]{babel}
\usepackage[T1]{fontenc}
\usepackage[utf8]{inputenc}
\usepackage{amsmath,amssymb,mathrsfs}
\usepackage{amscd}
\usepackage{tikz}
\usepackage[colorlinks=true,linkcolor= blue, citecolor=blue]{hyperref}

\usepackage{a4wide}

\numberwithin{equation}{section}
\newtheorem{theo}{Theorem}[section]
\newtheorem{cor}{Corollary}[section]

\newtheorem{prop}{Proposition}[section]
\newtheorem{lem}{Lemma}[section]
\newtheorem{de}{Definition}[section]
\newtheorem{ass}{Assumption}[section]
\newtheorem{rem}{Remark}[section]
\numberwithin{figure}{section}

\date{}

\newcommand{\bz}{{\mathbb{Z}}}
\newcommand{\br}{{\mathbb{R}}}
\newcommand{\bc}{{\mathbb{C}}}

\newcommand{\sinf}{{\textbf{\textup{S}}_\infty}}

\newcommand{\lao}{{H_0}}
\newcommand{\la}{{H_V}}

\newcommand{\im}{\operatorname{Im}}
\newcommand{\re}{\operatorname{Re}}

\newenvironment{proof}
	{Proof.}
	{\hfill $\square$\vskip 8pt}

\newcommand{\R}{\mathbb{R}}

\newcommand{\C}{\mathbb{C}}

\newcommand{\e}{\mathrm{e}}

\newcommand{\bra}{\langle}
\newcommand{\ket}{\rangle}

\title{On non-selfadjoint operators with finite discrete spectrum}

\author{Olivier {\sc Bourget}, Diomba {\sc Sambou}, and Amal {\sc Taarabt}}
\begin{document}

\maketitle

Facultad de Matem\'aticas, Pontificia Universidad Cat\'olica de Chile, Vicu\~na Mackenna 4860, Santiago de Chile 

\bigskip

E-mails: {\it bourget@mat.uc.cl, disambou@mat.uc.cl, amtaarabt@mat.uc.cl}

\begin{abstract}
We consider some compact non-selfadjoint perturbations of fibered one-dimensional discrete Schr\"odinger operators. We show that the 
perturbed operator exhibits finite discrete spectrum under suitable\- regularity conditions.
\end{abstract}


\bigskip

\tableofcontents

\section{Introduction}\label{s1}

The spectral theory of non-selfadjoint perturbations of selfadjoint operators has made significant advances in the last decade. For an historical panorama on the relationhips between non-selfadjoint operators and quantum mechanics, we refer the reader to e.g. \cite{bgsz} and references therein. The distributional properties of the discrete spectrum have been one of the main issues considered in this field, among the many results obtained so far.

In the present note, we keep going on with the spectral analysis of compact non-selfadjoint perturbations of the discrete Schr\"odinger operator. In \cite{bst}, we have proved that in dimension 1 and under adequate regularity conditions, the discrete spectrum of the perturbed operator remains finite. We have also exhibited some Limiting Absorption Principle, thus completing some previous results obtained by \cite{ego1}, \cite{ego2} in the framework of Jacobi matrices. Presently, we show that these results can actually be adapted to consider non-selfadjoint perturbations of some fibered version of the one dimensional Schr\"odinger operator.

Following the methodology developed in \cite{bst}, the paper is organized as follows. The model and the results are introduced in Section 2. Theorem \ref{t:tda} is proved in Section 3.1, through an analysis far from the thresholds involving complex scaling arguments. In Section 3.2, an analysis of the resonances located in a neighbourhood of the thresholds (Theorem \ref{t1s}) allows to conclude the proof of Theorem \ref{t1s+}.

\medskip

\noindent
{\bf Notations and basic concepts.} 
Throughout this paper, we adopt the notations of \cite{bst}. ${\mathbb Z}$, ${\mathbb Z}_+$ and ${\mathbb N}$ denote respectively the sets of integral numbers, 
non-negative and positive integral numbers. For $\gamma \geq 0$, we define the weighted Hilbert spaces
$$
\ell_{\pm \gamma}^{2}({\mathbb Z}) := \big\lbrace x \in {\mathbb C}^{\mathbb Z} : \sum_{n \in {\mathbb Z}} {\rm e}^{\pm \gamma \vert n \vert} |x(n)|^2 < \infty \big\rbrace .
$$
In particular, $\ell^{2}({\mathbb Z}) = \ell_{0}^{2}({\mathbb Z})$ and one has the inclusions $\ell_{\gamma}^{2}({\mathbb Z}) \subset \ell^{2}({\mathbb Z}) \subset \ell_{- \gamma}^{2}({\mathbb Z})$. For $\gamma > 0$, one defines the multiplication operators $W_\gamma : \ell_\gamma^2({\mathbb Z}) \rightarrow \ell^{2}({\mathbb Z})$ by $\left( W_\gamma x \right)(n) := {\rm e}^{(\gamma/2) \vert n \vert} x(n)$, and $W_{- \gamma}: \ell^2({\mathbb Z}) \rightarrow \ell_{- \gamma}^2({\mathbb Z}) $ by $\left( W_{- \gamma} x \right)(n) := {\rm e}^{- (\gamma/2) \vert n \vert} x(n)$. We denote by $(\delta_n)_{n\in {\mathbb Z}}$ the canonical orthonormal basis of $\ell^{2}({\mathbb Z})$. 

The discrete Fourier transform ${\mathcal F} : \ell^2({\mathbb Z}) \rightarrow {\rm L}^2({\mathbb T})$, where ${\mathbb T} := {\mathbb R} /2\pi {\mathbb Z}$, 
is defined for any $x \in \ell^2({\mathbb Z})$ and $f \in {\rm L}^2({\mathbb T})$ by
\begin{equation}\label{eq1,3}
({\mathcal F} x)(\vartheta) := \frac{1}{\sqrt{2\pi}} \sum_{n \, \in \, {\mathbb Z}} e^{-in\vartheta} x(n), \quad
\big( {\mathcal F}^{-1} f \big)(n) := \frac{1}{\sqrt{2\pi}} \int_{{\mathbb T}} e^{in\vartheta} f(\vartheta) d\vartheta.
\end{equation}
The operator ${\mathcal F}$ is unitary. For any bounded (resp. selfadjoint) operator $X$ acting on $\ell^2({\mathbb Z})$, we define the bounded (resp. selfadjoint) 
operator $\widehat X$ acting on ${\rm L}^2({\mathbb T})$ by
\begin{equation}\label{eq:mr2}
\widehat X := {\mathcal F} X {\mathcal F}^{-1}.
\end{equation}
More generally, if ${\mathscr H}$ is an auxiliary Hilbert space, for any bounded (resp. selfadjoint) operator $Y$ acting on $\ell^2({\mathbb Z})\otimes {\mathscr H}$, we define the bounded (resp. selfadjoint) operator $\widehat Y$ acting on ${\rm L}^2({\mathbb T})\otimes {\mathscr H}$ by
\begin{equation}
\widehat Y := ({\mathcal F} \otimes I) Y ({\mathcal F}^{-1} \otimes I).
\end{equation}

If ${\mathscr H}$ is a separable Hilbert space, $\mathcal{B}({\mathscr H})$ and ${\rm GL}({\mathscr H})$ denotes the algebras of bounded linear operators and boundedly invertible linear operators acting on ${\mathscr H}$. $\sinf({\mathscr H})$ stands for the ideal of compact operators. For any operator $H \in \mathcal{B}({\mathscr H})$, we denote its spectrum by $\sigma (H)$, its resolvent set by $\rho (H)$, the set of its eigenvalues by ${\mathscr E}_{\rm p} (H)$. We also define its point spectrum as the closure of the set of its eigenvalues and write it $\sigma_{\rm pp} (H)=\overline{{\mathscr E}_{\rm p} (H)}$. Since this article deals with non-selfadjoint (bounded) operators, it is also convenient to clarify the different notions of spectra we use. Let $T$ be a closed linear operator acting on a Hilbert space $\mathscr{H}$, and $z$ be an isolated point of $\sigma(T)$. If $\gamma$ is a small contour positively oriented containing $z$ as the only point of $\sigma(T)$, the Riesz projection $P_z$ associated to $z$ is defined by 
$$
P_z := \frac{1}{2i\pi} \oint_{\gamma} (T - \zeta)^{-1} d\zeta .
$$
The algebraic multiplicity of $z$ is then defined by 
\begin{equation}\label{eq1,9}
\textup{m}(z) := \text{rank} (P_z),
\end{equation}
and when it is finite, the point $z$ is called a discrete eigenvalue of the operator $T$. Note that one has the inequality $\mathrm{m} \, (z) \geq \mathrm{dim} \, \big( \text{Ker} (T - z) \big)$, which is the geometric multiplicity of $z$. The equality holds if $T$ is normal (see e.g. \cite{kato}). So, one defines the discrete spectrum of $T$ as
\begin{equation}\label{eq1,10}
\sigma_{\text{disc}}(T) := \big\lbrace z \in \sigma(T) : z \hspace*{0.1cm} 
\textup{is a discrete eigenvalue of $T$} \big\rbrace.
\end{equation}
A closed linear operator is said to be of Fredholm if it has a closed range and both its kernel and cokernel are finite-dimensional. We define the essential spectrum of $T$ as
\begin{equation}\label{eq1,11}
\sigma_{\text{ess}}(T) := \big\lbrace z \in \bc : \textup{$T - z$ \textup{is not a 
Fredholm operator}} \big\rbrace.
\end{equation} 

For $\Omega \subseteq {\mathbb C}$ an open domain and ${\mathbb B}$ a Banach space, ${\rm Hol}(\Omega,\mathbb{B})$ denotes the set of holomorphic functions from $\Omega$ with values in ${\mathbb B}$. For two  subsets $\Delta_1$ and $\Delta_2$ of ${\mathbb R}$, we denote as a subset of ${\mathbb C}$, $\Delta_1 + i \Delta_2 := \big\lbrace z \in {\mathbb C} : \re(z) \in \Delta_1 \, {\rm and} \, \im(z) \in \Delta_2 \big\rbrace$. For $R > 0$ and $\zeta_0 \in {\mathbb C}$, we set $D_R(\zeta_0) := \big\lbrace z \in \bc : |z - \zeta_0| < R \big\rbrace$ and $D_R^\ast(\zeta_0) := D_R(\zeta_0) \setminus \lbrace \zeta_0 \rbrace$. By $0 < \vert \eta \vert <\!\!<1$, we mean that $\eta \in {\mathbb C} \setminus \{ 0 \}$ is sufficiently close to $0$.

\section{Model and main results}\label{secd}

Let $\mathscr{H}$ be a separable Hilbert space and $\ell^2(\bz,\mathscr{H})$ the Hilbert space endowed with the scalar product 
$\displaystyle \langle \phi,\psi \rangle := \displaystyle \sum_{n \in \bz} \langle \phi(n),\psi(n) \rangle_\mathscr{H} .
$

\noindent Let $H_0$ be the bounded selfadjoint operator defined on $\ell^2(\bz,\mathscr{H})$ by
\begin{equation}\label{eq:g+1}
(H_0 \phi)(n) := 2 \phi(n) - \phi(n+1) - \phi(n-1) .
\end{equation}
Since $\ell^2(\bz,\mathscr{H}) \cong \ell^2(\bz)\otimes \mathscr{H}$, we will also look at $H_0$ as an operator acting on $\ell^2(\bz)\otimes \mathscr{H}$ and write: $H_0 = L_0 \otimes I$, where $L_0$ is the one-dimensional Schr\"odinger operator acting on $\ell^2(\bz)$ as
\begin{equation}\label{eq:g+2}
(L_0 x)(n) := 2 x(n) - x(n+1) - x(n-1) .
\end{equation}
In particular, $\sigma (H_0) = \sigma (L_0) = [0,4]$. The operators $L_0$ and $H_0$ are purely absolutely continuous, and so
\begin{equation}\label{eq:g+4}
\sigma (H_0) = \sigma_{\textup{ac}} (H_0) = \sigma_{\textup{ess}} (H_0) = [0,4].
\end{equation}

Using Fourier transform and the fact that ${\rm L}^2({\mathbb T}; {\mathscr H}) \cong {\rm L}^2({\mathbb T})\otimes {\mathscr H}$, we will also look at $\widehat \lao$ as an operator  acting on ${\rm L}^2({\mathbb T})\otimes {\mathscr H}$ and write: $\widehat \lao = \widehat L_0 \otimes I$, where $\widehat L_0$ is the multiplication operator on ${\rm L}^2({\mathbb T})$ by the function $f$:
$$
f(\alpha) := 2 - 2\cos \alpha = \left( 4 \sin^2 \frac{\alpha}{2} \right), \quad \alpha \in {\mathbb T}.
$$
In the sequel, the operator $\widehat L_0$ is identified with $f$.

For a potential $V \in \mathcal{B} \big( \ell^2(\bz,\mathscr{H})  \big)$, we denote the perturbed operator $H_V$ on $\ell^2(\bz,\mathscr{H})$ by
\begin{equation}\label{eq:g+8}
(H_V \phi)(n)  := (H_0 \phi)(n)  + ( V \phi)(n)\quad , \quad n\in \bz .
\end{equation}
Correspondingly, $\widehat{H_V}$ is defined on ${\rm L}^2({\mathbb T}; {\mathscr H})$ by: $\widehat{H_V} = \widehat{H_0}+\widehat{V}$.

\subsection{Spectrum and Limiting Absorption Principles}\label{sec:2,1}

Following \cite{bst}, we start by recalling the concept of complex scaling/analytic distortion w.r.t. a selfadjoint operator $A$ (see Section \ref{se} for more details and examples).

\begin{de}\label{dcl} 
Let ${\mathscr H}$ be a Hilbert space and $R > 0$. Let $A$ be a selfadjoint operator defined on ${\mathscr H}$. An 
operator $B \in \mathcal{B}({\mathscr H})$ belongs to the class ${\mathcal A}_R(A)$ if the map $\theta \mapsto {\rm e}^{i \theta A} B {\rm e}^{-i \theta A}$, 
defined for $\theta \in \br$, has an extension lying in ${\rm Hol} \big( D_R(0),\mathcal{B}({\mathscr H}) \big)$. In this case, we write $B \in {\mathcal A}_R(A)$. The collection of bounded operators for which a complex scaling w.r.t. $A$ can be performed is simply denoted by: ${\mathcal A}(A) := \displaystyle\bigcup_{R>0} {\mathcal A}_R(A)$. 
\end{de}
The main properties of the classes ${\mathcal A}(A)$ are listed in \cite[Section 6]{bst}, and we will frequently refer to them (see Section \ref{sa}).

Given the operator $H_0$ introduced previously, we define the auxiliary selfadjoint operator ${\bf A}_0$ acting on $\ell^2(\bz,\mathscr{H})$ by:
\begin{equation}\label{A0}
{\bf A}_0 := A_0 \otimes I,
\end{equation}
where $A_0 := {\mathcal F}^{-1} \widehat A_0 {\mathcal F}$ and the operator $\widehat A_0$ is the unique selfadjoint extension of the symmetric operator defined on $C^{\infty}({\mathbb T})$ by
$\widehat A_0 := \sin \vartheta (-i\partial_\vartheta) + (-i\partial_\vartheta) \sin \vartheta.$
The operators ${\bf A}_0$ and $A_0$ are respectively the conjugate operator of $H_0$ and $L_0$ in the sense of the Mourre theory.

\begin{rem}\label{r:sae}
Some examples of operators which belong to $\mathcal{A}({\bf A}_0 )$ are
\begin{enumerate}
\item[(i)] $V = |\psi \rangle \langle \varphi | \otimes B$ where $\varphi$ and $\psi$ are analytic vectors for $A_0$ and $B \in \mathcal{B}({\mathscr H})$. For 
instance, the canonical orthonormal basis $(\delta_n)_{n \in \bz}$ of $\ell^2(\bz)$ is a family of analytic vectors for $A_0$. In particular, for each $n \in \bz$ 
and $u \in \mathscr{H}$, $\delta_n \otimes u$ is an analytic vector for ${\bf A}_0$ (see Section \ref{se}).
\item[(ii)] $V$ satisfying Assumption \ref{ass:g+1} with $\Gamma_j = \mu_j I$, $j = 1$, $2$, $\mu_j \in \bc$ (see Proposition \ref{pe3}).
\end{enumerate}
\end{rem}

If the perturbation $V \in \sinf \big( \ell^2(\bz,\mathscr{H}) \big)$ is compact, it follows from the Weyl criterion on the invariance of the essential spectrum under compact perturbations and from \cite[Theorem 2.1, p. 373]{goh}, that one has the disjoint union
$\sigma(H_V) = \sigma_{{\rm ess}}(H_V) \bigsqcup  \sigma_{{\rm disc}}(H_V)$, where $\sigma_{{\rm ess}} (H_V) = [0,4]$. Furthermore, 
the only possible limit points of $\sigma_{{\rm disc}}(H_V)$ are contained in $\sigma_{{\rm ess}} (H_V)$. Under additional regularity conditions on $V$, the next 
theorems give more information about the distribution of $\sigma_{{\rm disc}}(H_V)$ near $\sigma_{{\rm ess}} (H_V)$, and that of ${\mathscr E}_{\rm p} (H_V)$ 
inside $\sigma_{{\rm ess}} (H_V)$.

\begin{theo}\label{t:tda} 
Let $V \in \sinf \big( \ell^2({\mathbb Z},\mathscr H) \big) \cap {\mathcal A}_R({\bf A}_0)$ for some $R > 0$. Then
\begin{enumerate}
\item The possible limit points of $\sigma_{\textup{disc}} (H_V)$ belong to the spectral thresholds $\lbrace 0,4 \rbrace$.
\item There exists a discrete subset ${\mathcal D} \subset (0,4)$ 
whose only possible limits points belong to $\{0,4\}$ and for which the following holds: given any relatively compact interval $\Delta_0$, 
$\overline{\Delta_0} \subset (0,4)\setminus {\mathcal D}$, there exist $\epsilon_0^{\pm} >0$ such that for any vectors $\varphi$ and $\psi$ analytic w.r.t. 
${\bf A}_0$,
\begin{gather*}
\sup_{z\in \Delta_0 \pm i(0,\epsilon_0^\pm)} |\langle \varphi, (z-H_V)^{-1} \psi \rangle | < \infty.
\end{gather*}
\end{enumerate}
\end{theo}

\begin{rem} If $H_V$ is selfadjoint, ${\mathcal D}$ coincides with the set of its embedded eigenvalues. If not, i.e. if $H_V \neq H_V^\ast$, we expect at least that the embedded eigenvalues belong to ${\mathcal D}$. We refer to Section \ref{p:scs} for a proof of Theorem \ref{t:tda}. 
\end{rem}

\subsection{Resonances}\label{ssrs}

The reader may have noted that Theorem \ref{t:tda} does not give any information about the distribution of ${\mathscr E}_{\rm p} (H_V)$ around the spectral thresholds $\lbrace 0,4 \rbrace$. This is a nontrivial problem ; under Assumption \ref{ass:g+1} below, we give an answer by means of resonances techniques and characteristic values methods (see e.g. \cite{bo}). First, we fix some notations and definitions.

Consider an orthonormal basis $(e_j)_{j \in {\mathbb Z}_+}$ of the Hilbert space ${\mathscr H}$. It follows that $(\delta_n \otimes e_j)_{(n,j)\in {\mathbb Z}\times {\mathbb Z}_+}$ is an orthonormal basis of $\ell^2(\bz,\mathscr{H})$. For each $j\in {\mathbb Z}_+$, we define the subspace ${\mathscr H}_j =$ span $\{ x \otimes e_j; x\in \ell^2(\bz)\}$ together with its corresponding orthogonal projection $P_j := I  \otimes |e_j \ket \bra e_j |$. Of course,
$$
\ell^2(\bz,\mathscr{H}) \cong \ell^2(\bz) \otimes  \mathscr{H} = \bigoplus_{j \ge 0} \mathscr{H}_j.
$$

We observe that for each $j \ge 0$, $\mathscr{H}_j$ is $H_0$-invariant and that $H_0$ rewrites:
\begin{equation}\label{eq:g+3}
H_0 = \bigoplus_{j \in {\mathbb Z}_+} P_j H_0 P_j = \bigoplus_{j \in {\mathbb Z}_+} L_0 \otimes |e_j \ket \bra e_j | = L_0 \otimes I .
\end{equation}


If $W = \big( w(n,m) \big)_{(n,m) \in \bz^2}$ is a matrix operator with coefficients $w(n,m) \in \mathcal{B}(\mathscr{H})$, then
\begin{equation}\label{eq:g+9}
(W \phi)(n) = \sum_{m \in \bz} w(n,m) \phi(m), \quad n \in \bz.
\end{equation}
In the orthonormal basis $(e_j)_{j \ge 0}$, for each $(n,m) \in \bz^2$, the operator $w(n,m)$ has the following matrix representation (with an abuse of notation)
\begin{equation}\label{eq:g+9+}
w(n,m) = \big( w_{jk}(n,m) \big)_{j,k \ge 0}, \qquad w_{jk}(n,m) := \langle e_j,w(n,m) e_k \rangle_\mathscr{H}.
\end{equation}

\begin{rem}\label{r:spd}
\mbox{}
\begin{itemize}
\item[(i)] If $\mathscr{H} = \bc^d$ for $d \ge 1$, then $w(n,m) \in \mathcal{M}_d(\bc)$. If $d = 1$ (i.e. $\mathscr{H} = \bc$), $W = \big( w(n,m) \big)_{(n,m) \in \bz^2}$ coincides with the matrix representation of $W$ in the canonical orthonormal basis of $\ell^2(\bz)$.
\item[(ii)] Another representation of $W$
is to write $W = \sum_{n,m} W_{n,m}$ where for each $(n,m) \in \bz^2$ fixed, $W_{n,m} = \big( w(\ell,k) \delta_{nm}\big)_{(\ell,k) \in \bz^2}$ corresponds to the matrix operator whose coefficients are
$$
w_{nm}(\ell,k) 
= \begin{cases} 
0 & \text{if} \quad (\ell,k) \neq (n,m), \\ 
w(n,m) & \text{if} \quad (\ell,k) = (n,m).
\end{cases}
$$
Therefore, for any $\phi \in \ell^2(\bz,\mathscr{H})$, we have $W_{n,m} \phi = \delta_n \otimes w(n,m) \phi(m)$, which implies that 
\begin{equation}
\label{eq:g+10}
W_{n,m} = | \delta_n \rangle \langle \delta_m | \otimes w(n,m).
\end{equation}
Thus, in $\ell^2(\bz) \otimes  \mathscr{H}$, $W$ has a canonical representation given by
\begin{equation}
\label{eq:g+11}
W = \sum_{n,m} | \delta_n \rangle \langle \delta_m | \otimes w(n,m).
\end{equation}
See also Remark \ref{rfin} for sufficient conditions for the compactness of the operators $W_{n,m}$ and $W$, using the representations \eqref{eq:g+10}
and \eqref{eq:g+11}.
\end{itemize}
\end{rem}

Let ${\bf J}$ be the selfadjoint unitary operator defined on $\ell^2(\bz,\mathscr H)$ by 
\begin{equation}\label{eq:s5,26}
{\bf J} := J \otimes I \quad {\rm with} \quad (J \varphi)(n) := (-1)^{\vert n \vert} \varphi(n).
\end{equation}
Note that the operator $J$ commutes with any multiplication operator acting on $\ell^2(\bz)$.
Our general assumption on $V$ is the following:

\begin{ass}\label{ass:g+1}
\mbox{}
\begin{itemize}
\item[(i)] $V$ is (non)-selfadjoint of the form $V = (\Gamma_1 \otimes \Lambda_1) W (\Gamma_2 \otimes \Lambda_2)$ where:
\begin{itemize}
\item $W$ is given by  \eqref{eq:g+9},
\item $\exists$ $\gamma > 0$ such that $\Gamma_j \in {\mathcal B}\big( \ell^2(\bz) \big)$, $j = 1$, $2$, commute with the operators $W_{-\gamma}$ and $J$,
\item $\Lambda_1 \in {\mathcal B}(\mathscr H)$, $\Lambda_2 \in \sinf(\mathscr H)$ and $\Lambda_2 \Lambda_1$ is of finite rank.
\end{itemize}
\item[(ii)] $\sup_{(n,m) \in \bz^2} \| w(n,m) \|_\mathscr{H} \, {\rm e}^{\gamma( |n| + |m|)} \le {\rm C}$
for each $(n,m) \in \bz^2$ and for some constant $C>0$, $\gamma$ being the constant introduced above.
\end{itemize}
\end{ass}

\begin{rem}\label{r:spd+}
\mbox{}
\begin{itemize}
\item[(i)] Of course if ${\rm dim}(\mathscr H) < \infty$, the last condition in Assumption \ref{ass:g+1} (i) on the $\Lambda_j$, $j = 1$, $2$
holds trivially. Moreover, $V = W$ coincides with the case where $\Gamma_j \otimes \Lambda_j = I$.
\item[(ii)] If ${\rm dim}(\mathscr H) = \infty$, the compact operator $\Lambda_2$ plays a regularization role in the component $\mathscr H$ of the space
$\ell^2(\bz) \otimes \mathscr H$, which is crucial to define the resonances.
\item[(iii)] Assumption \ref{ass:g+1} (ii) holds for $\beta > 2$ and
$$ 
\big| w_{jk}(n,m) \big| \le {\rm } \big\langle (j,k) \big\rangle^{-\beta} {\rm e}^{-\gamma( |n| + |m|)}, (j,k) \in \bz_+^2, \gamma > 0, (n,m) \in \bz^2.
$$
\item[(iv)] A perturbation $V$ which satisfies Assumption \ref{ass:g+1} is compact (see Lemma \ref{l:sdecU}).
\end{itemize}
\end{rem}

For our purpose in the sequel, let us recall that under Assumption \ref{ass:g+1}, the resonances of the operator $H_V$ near the spectral thresholds 
$\lbrace 0,4 \rbrace$ are defined as poles of the meromorphic extension of the resolvent $(H_V - z)^{-1}$ in some Banach weighted spaces. Moreover, 
they are parametrized respectively near $0$ and $4$ by 
\begin{equation}\label{eq:prs}
z_0(\lambda) := \lambda^2 \quad {\rm and} \quad z_4(\lambda) := 4 - \lambda^2,
\end{equation}
and are defined in some two-sheets Riemann surfaces. In particular, the discrete and the embedded eigenvalues of $H_V$ near $\lbrace 0,4 \rbrace$
are resonances. One refers to Definitions \ref{d:s5,1}, \ref{d:s5,2} and Section \ref{p:sr} for more details. 

\begin{rem}
\mbox{}
\begin{itemize}
\item[(i)] The resonances can be defined by meromorphic extension of the weighted resolvent (see Section \ref{ssrs}) or by complex dilation (see Section \ref{sec:2,1}).
When the resonances are discrete eigenvalues, these two definitions coincide. If not, the problem is open. One refers for instance to the article \cite{helmar} by Helffer and
Martinez, where it is proved, for perturbations of the Schr\"odinger operator in a semi-classical regime, that under some general conditions, such definitions coincide.
\item[(ii)] Under Assumption \ref{ass:g+1}, the resonances are defined in pointed neighborhoods of $\{ 0,4 \}$, i.e. pointed at the thresholds $\{ 0,4 \}$. The problem 
of knowing if $\{ 0,4 \}$ are resonances or not is open.
\item[(iii)] The assumption "$\Lambda_2 \Lambda_1$ is of finite rank" is very restrictive in the sense that we want to investigate here only finiteness properties of the discrete
spectrum. However, it is possible to study the accumulation of eigenvalues (or resonances) at $\{ 0,4 \}$ without this assumption.
\end{itemize}
\end{rem}

Our second main result is the following:

\begin{theo}\label{t1s} 
Let $V$ satisfy Assumption \ref{ass:g+1}. Then, for any $0 < r <\!\!<1$, $\mu \in \lbrace 0,4 \rbrace$, there is no
$\lambda \in D_r^\ast (0)$ such that $z_\mu(\lambda)$ is a resonance of $H_V$.
\end{theo}

Theorem \ref{t1s} just says that the operator $H_V$ has no resonances in a punctured neighborhood of $0$ and $4$ in the two-sheets Riemann surfaces
where they are defined, see Fig. \ref{fig 1} below for a graphic illustration. The proof of Theorem \ref{t1s} is postponed to Section \ref{p:sr}.

\begin{figure}[h]
\begin{center}
\tikzstyle{+grisEncadre}=[fill=gray!60]
\tikzstyle{blancEncadre}=[fill=white!100]
\tikzstyle{grisEncadre}=[fill=gray!20]
\begin{tikzpicture}[scale=1.2]

\draw [grisEncadre] (0,0) -- (180:1.58) arc (180:360:1.58) -- cycle;
\draw [blancEncadre] (0,0) -- (0:1.58) arc (0:180:1.58) -- cycle;

\draw [->] [thick] (-2.5,0) -- (2.3,0);
\draw (2.3,0) node[right] {\small{$\re(\lambda)$}};
\draw [->] [thick] (0,-2.5) -- (0,2.7);
\draw (0.05,2.7) node[right] {\small{$\im(\lambda)$}};
\draw (0,0) -- (1.31,0.9);
\draw (0.8,0.55) node[above] {\tiny{$r$}};

\node at (-1.4,-1.2) {\tiny{$\times$}};

\node at (0.9,-1.4) {\tiny{$\times$}};
\node at (1.8,-0.2) {\tiny{$\times$}};
\node at (1.4,-1.2) {\tiny{$\times$}};

\node at (0.9,1.4) {\tiny{$\times$}};
\node at (1.4,1.2) {\tiny{$\times$}};

\node at (-1.7,0.4) {\tiny{$\times$}};
\node at (-1.4,1.2) {\tiny{$\times$}};

\node at (-0.5,1.7) {\tiny{$\times$}};

\node at (-0.2,1.7) {\tiny{$\times$}};

\node at (2,2.3) {\small{$\textup{\textbf{Absence of resonances}}$}};
\draw [->] [dotted] (1.8,2.15) -- (-0.5,1);
\draw [->] [dotted] (1.8,2.15) -- (1,-0.5);

\node at (-4.5,0.5) {\small{$\textup{Corresponds to the \textbf{physical plane}}$}};

\draw [->] [dotted] (-2.1,0.5) -- (-1,0.5);

\node at (-4.67,-0.5) {\small{$\textup{Corresponds to the \textbf{nonphysical}}$}};
\node at (-4.55,-0.8) {\small{$\textup{\textbf{plane}}$}};
\draw [->] [dotted] (-1.8,-0.5) -- (-1,-0.5);

\end{tikzpicture}

\caption{\textup{\textbf{Resonances near $z_\mu(\lambda)$ in variable $\lambda$:} Thanks to Theorem \ref{t1s}, $H_V$ has 
no resonance $z_\mu(\lambda)$ in $\big\lbrace \lambda : 0 < \vert \lambda \vert < r \big\rbrace$ for $\mu \in \lbrace 0,4 \rbrace$ and $r$ small enough.}}\label{fig 1}
\end{center}
\end{figure}

\medskip

According to Remark  \ref{r:sae} (ii) and Remark \ref{r:spd+} (iv), perturbations $V$ satisfying Assumption \ref{ass:g+1} with $\Gamma_j \otimes \Lambda_j = I$, 
$j = 1$, $2$, verify $V \in \sinf \big( \ell^2({\mathbb Z},\mathscr H) \big) \cap {\mathcal A}_{R_\gamma} ({\bf A}_0)$ for some $R_\gamma > 0$. Thus, one deduces 
from Theorems \ref{t:tda} and \ref{t1s} the following result:

\begin{theo}\label{t1s+} 
Let $V$ satisfy Assumption \ref{ass:g+1} with $\Gamma_j = \mu_j I$, $j = 1$, $2$, $\mu_j \in \bc$. Then $\sigma_{\textup{disc}} (H_V)$ has no limit points 
in $[0,4]$, and hence is finite.
\end{theo}

It follows from Theorem \ref{t1s} and the usual complex scaling arguments \cite{sig} the following:

\begin{cor}\label{c1s} 
Assume that the perturbation $V$ is selfadjoint and satisfies Assumption \ref{ass:g+1} with $\Gamma_j = \mu_j I$, $j = 1$, $2$, $\mu_j \in \bc$. Then:
\begin{itemize}
\item $\sigma_{\rm {ess}} (H_V)= [0,4]$ and $\sigma_{{\rm disc}} (H_V)$ is finite.
\item There is at most a finite numbers of eigenvalues embedded in $[0,4]$, each of these eigenvalues having finite multiplicity.
\item The singular continuous spectrum $\sigma_{{\rm sc}} (H_V) = \emptyset$ and the following LAP holds: given any relatively compact interval
$\Delta_0 \subset (0,4) \setminus {\mathscr E}_{\rm p}(H_V)$, there exist $\epsilon_0^{\pm} >0$ such that for any vectors $\varphi$ and $\psi$ analytic 
w.r.t. ${\bf A}_0$,
\begin{gather*}
\sup_{z\in \Delta_0 \pm i(0,\epsilon_0^\pm)} |\langle \varphi, (z-H_V)^{-1} \psi \rangle | < \infty.
\end{gather*}
\end{itemize}
\end{cor}

\begin{rem} Setting ${\mathscr H}={\mathbb C}$ in Theorems \ref{t:tda}, \ref{t1s}, \ref{t1s+} and Corollary \ref{c1s}, we recover Theorems 2.2, 2.3, 2.4, 2.5 and Corollary 2.1 of \cite{bst}.
\end{rem}

\section{Proofs of the main results}\label{p:s}

\subsection{Complex scaling}\label{p:scs}

In this section, we take advantage of the complex scaling techniques developed in \cite{bst} to study $\sigma (H_V)$ for compact perturbations $V \in {\mathcal A}({\bf A}_0)$. Since the operators $\la$ and $\widehat \la$ are unitarily equivalent, we focus our attention on the analysis on the latter. Note that $H_V \in {\mathcal A}_R({\bf A}_0)$ for some $R>0$ if and only if ${\widehat H}_V\in {\mathcal A}_R (\widehat {\bf A}_0)$.

\subsubsection{Complex scaling for $H_0$}

We describe the complex scaling process for the unperturbed operator $\widehat \lao$. For complementary references, see e.g. \cite{hissig}, \cite{RS4} and \cite{sig}.

First, we observe that $f= T\circ \cos$, where $T:\C \rightarrow \C$, $T(z) := 2(1 - z)$. The map $T$ is bijective and maps $[-1,1]$ onto $[0,4]$. The points $T(-1)=4$ and $T(1)=0$ are the thresholds of $\lao$ and $\widehat{\lao}$.

We also have that for any $\theta\in\R,$
$$
\e^{i\theta {\widehat {\bf A}}_0} = \e^{i\theta {\widehat A}_0} \otimes I .
$$ 
It follows from \cite[Section 3.1]{bst}, that for any $\theta\in\R,$
\begin{equation}\label{eq6,6+}
\e^{i\theta {\widehat {\bf A}}_0} \widehat \lao \e^{-i\theta {\widehat {\bf A}}_0} = 
\e^{i\theta \widehat A_0} \widehat L_0 \e^{-i\theta \widehat A_0} \otimes I = (G_{\theta} ( {\widehat L}_0)) \otimes I,
\end{equation} 
where for $\theta \in \R$, the function $G_\theta$ is defined on $[0,4]$ by $G_\theta := T \circ F_{\theta} \circ T^{-1}$ with
\begin{equation}\label{eq6,7}
F_\theta(\lambda) := \frac{\lambda - {\rm th}(2\theta)}{1 - \lambda {\rm th}(2\theta)}, \qquad \lambda \in [-1,1].
\end{equation}

\begin{rem} In \eqref{eq6,7}, the denominator does not vanish since $\big\vert {\rm th}(2\theta) \lambda \big\vert < 1$ for $\lambda \in [-1,1]$.
\end{rem}

In order to perform our complex scaling argument, we need to precise the meaning of formula \eqref{eq6,6+} and \eqref{eq6,7}, for possibly complex values of the deformation parameter $\theta$. To this end, we observe that:
\begin{prop}\label{p6,1} Let $\mathbb{D} := D_1(0)$ denote the open unit disk of the complex plane $\C$. Then,
\begin{enumerate}
\item[(a)] For any $\lambda \in [-1,1]$, the map $\theta \mapsto F_\theta(\lambda)$ is holomorphic in $D_\frac{\pi}{4}(0)$.
\item[(b)] For $\theta \in {\mathbb C}$ such that $\vert \theta \vert < \frac{\pi}{8}$, the map $\lambda \mapsto F_\theta(\lambda)$ is a homographic transformation with 
$F_\theta^{-1} = F_{-\theta}$. In particular, for $\theta \in \R$, $F_\theta(\mathbb{D}) = \mathbb{D}$ and $F_\theta ( [-1,1] ) = [-1,1]$.
\item[(c)] For $\theta \in {\mathbb C}$ such that $0 < \vert \theta \vert < \frac{\pi}{8}$, the unique fixed points of $F_\theta$ are $\pm 1$.
\item[(d)] For $\theta_1$, $\theta_2 \in {\mathbb C}$ with $\vert \theta_1 \vert$, $\vert \theta_2 \vert < \frac{\pi}{8}$, we have that: $F_{\theta_1} \circ F_{\theta_2} 
= F_{\theta_1 + \theta_2}$.
\end{enumerate}
\end{prop}

From \eqref{eq6,6+} and Proposition \ref{p6,1}, it follows that:
\begin{prop}\label{p6,2}
The bounded operator valued-function
$$
\theta \mapsto \e^{i\theta {\widehat {\bf A}}_0} \widehat \lao \e^{-i\theta {\widehat {\bf A}}_0} = \e^{i\theta \widehat A_0} \widehat L_0 \e^{-i\theta \widehat A_0} \otimes I \in \mathcal{B} \big( {\rm L}^2({\mathbb T},\mathscr H) \big),
$$
admits an analytic extension from $\big( -\frac{\pi}{8},\frac{\pi}{8} \big)$ to $D_\frac{\pi}{8}(0)$, with extension given for $\theta \in D_\frac{\pi}{8}(0)$ 
by the operator-valued map $\theta \mapsto G_{\theta} (\widehat L_0) \otimes I$, where $G_{\theta} (\widehat L_0)$ is the multiplication operator by the function $G_{\theta} \circ f = T \circ F_{\theta} \circ \cos = G_{\theta} \circ T \circ \cos$. In the sequel, this extension is denoted $\widehat \lao(\theta)$.
\end{prop}

\begin{figure}[!h]
\begin{center}
\includegraphics[scale=0.75]{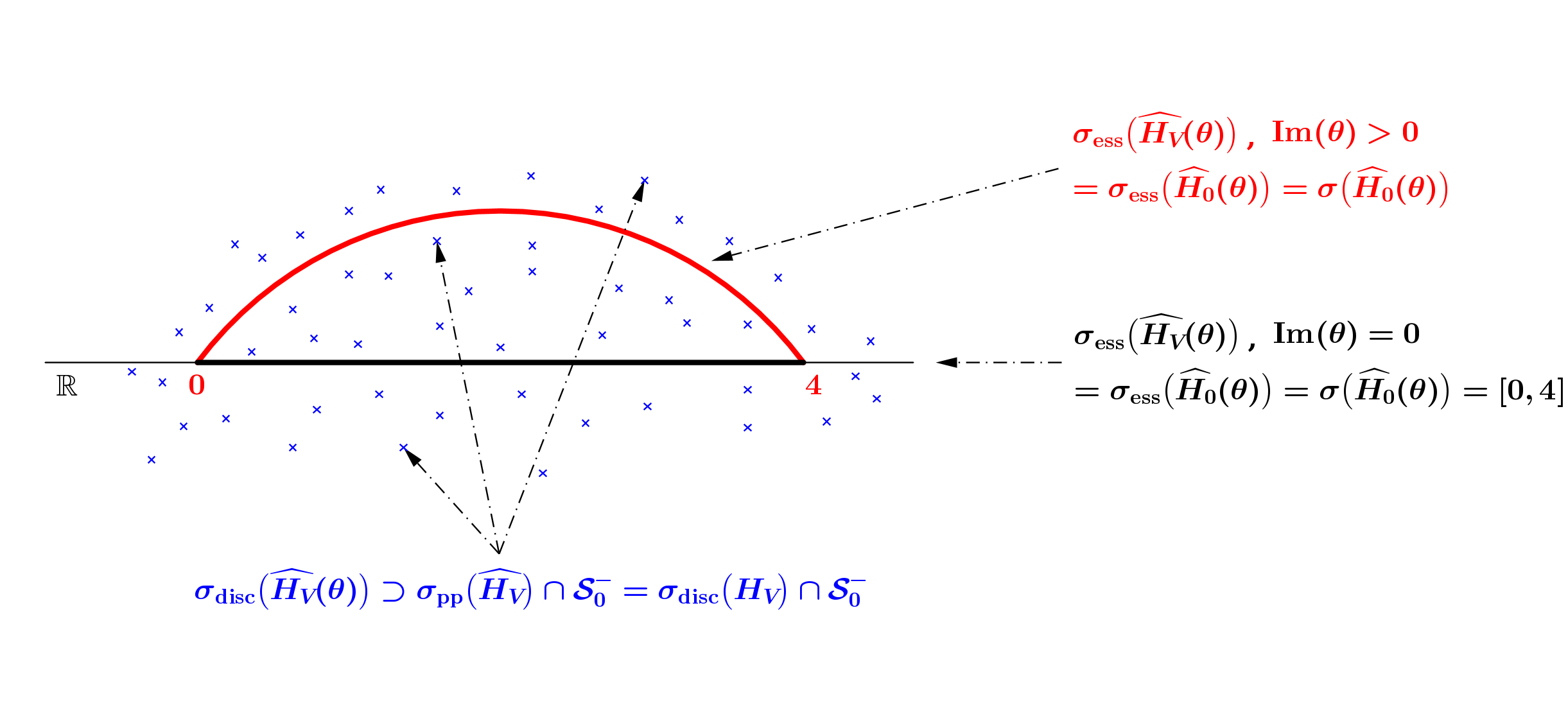}
\vspace*{-0.7cm}
\caption{Spectral structure of the operator $\widehat \la(\theta)$ for $\theta \in D_\frac{\pi}{8}(0)$ and $\im(\theta) \ge 0$.} \label{figd}
\end{center}
\end{figure}

Paraphrasing Proposition \ref{p6,2}, we have that $\widehat{H}_0 \in {\mathcal A}_{\frac{\pi}{8}}(\widehat{\bf A}_0)$ and for any $\theta \in D_\frac{\pi}{8}(0),$
$$
\widehat \lao(\theta) = G_{\theta} (\widehat L_0) \otimes I .
$$
Combining the continuous functional calculus, Proposition \ref{p6,2} and unitary equivalence properties, we get for $\theta \in D_\frac{\pi}{8}(0),$
\begin{equation*}
\sigma ( \widehat \lao(\theta) ) = \sigma ( G_\theta (\widehat \lao) ) = G_\theta ( \sigma (\widehat \lao) ) 
= G_\theta ( \sigma (\lao) ) = G_\theta ( [0,4] ).
\end{equation*}
Thus, for $\theta \in D_\frac{\pi}{8}(0)$, $\sigma \big( \widehat \lao(\theta) \big)$ is a smooth parametrized curve described by
$$
\sigma \big( \widehat \lao(\theta) \big) = \big\lbrace T\circ F_{\theta}(\lambda) = G_{\theta}\circ T (\lambda) : \lambda \in [-1,1] \big\rbrace.
$$
Quoting \cite[Proposition 3.3]{bst}, the following properties hold:

\begin{prop}\label{p6,3} 
Let $(\widehat \lao(\theta))_{\theta \in D_\frac{\pi}{8}(0)}$ be the family of bounded operators defined above. Then:
\begin{itemize}
\item[(i)] For $\theta_1$, $\theta_2 \in D_\frac{\pi}{8}(0)$ such that $\im(\theta_1) = \im(\theta_2)$, one has
$\sigma \big( \widehat \lao(\theta_1) \big) = \sigma \big( \widehat \lao(\theta_2) \big)$.
That is, the curve $\sigma \big( \widehat \lao(\theta) \big)$ does not depend on the choice of $\re(\theta)$.
\item[(ii)] For $\pm \im(\theta) > 0$, $\theta \in D_\frac{\pi}{8}(0)$, the curve $\sigma \big( \widehat \lao(\theta) \big)$ lies in ${\mathbb C}_\pm$.
\item[(iii)] Let $\theta \in D_\frac{\pi}{8}(0)$. If $\im(\theta) \neq 0$, the curve $\sigma \big( \widehat \lao(\theta) \big)$ is an arc of a circle containing the points $0$ and $4$. If $\im(\theta) = 0$, $\sigma \big( \widehat \lao(\theta) \big) = [0,4]$.
\end{itemize}
\end{prop}
We refer to Fig. \ref{figd} below for a graphic illustration.


\subsubsection{Complex scaling for $H_V$}

Now, we proceed to the complex scaling of the perturbation $\widehat V$ and the corresponding perturbed operator $\widehat \la := \widehat \lao + \widehat V$. 
For $\widehat V \in {\mathcal A}_R (\widehat {\bf A}_0)$, $R > 0$, and for all $\theta \in D_R(0)$, we write: $\widehat V(\theta) := \e^{i\theta \widehat {\bf A}_0} \widehat V \ \e^{-i\theta \widehat {\bf A}_0}$.

Quoting \cite[Lemma 5, Section XIII.5]{RS4}, we recall that:
\begin{lem}\label{ld1}
Let $\widehat V \in \sinf \big( {\rm L}^2({\mathbb T},\mathscr H) \big) \cap {\mathcal A}_R (\widehat {\bf A}_0)$, for some $R > 0$. Then, for all $\theta\in D_R(0)$, 
$\widehat V(\theta)$ is compact.
\end{lem}

Following Proposition \ref{p6,2}, we can define for $\widehat V \in {\mathcal A}_R (\widehat {\bf A}_0)$ and all $\theta \in D_{2R'}(0)$ with $2R' := \min \big(R,\frac{\pi}{8} \big),$
$$
\widehat \la(\theta) := \widehat \lao(\theta) + \widehat V(\theta) .
$$
By construction, $(\widehat \la(\theta))_{\theta \in D_{2R'}(0)}$ is a holomorphic family of bounded operators on $D_{2R'}(0)$. Also, the operator $\widehat{H_V}$ belongs to ${\mathcal A}_{2R'} (\widehat {\bf A}_0)$ and the map $\theta \mapsto \widehat{H_V} (\theta)$ actually coincides with the holomorphic extension of the map $\theta \mapsto \e^{i\theta \widehat A_0} \widehat \la \e^{-i\theta \widehat A_0}$ from the interval $(-2R', 2R')$ to $D_{2R'}(0)$. In addition, we have that:
\begin{prop}\label{pdr1}
Let $\widehat V \in {\mathcal A}_R (\widehat A_0)$, $R > 0$. Then, for any $\theta' \in \R$ such that $\vert \theta' \vert < R'$, we have
\begin{equation*}
\widehat \la(\theta+ \theta')=\e^{i \theta \widehat A_0} \widehat \la(\theta')\ \e^{-i \theta \widehat A_0},
\end{equation*}
for all $\theta\in D_{R'}(0)$.
\end{prop}
\noindent \begin{proof} Fix $\theta' \in \R$ with $\vert \theta' \vert <R'$ and observe that the following maps
\begin{align*}
\theta \longmapsto \e^{i\theta' \widehat A_0} \widehat \la(\theta) \e^{-i\theta' \widehat A_0} \quad {\rm and} \quad \theta \longmapsto \widehat \la(\theta+\theta'),
\end{align*}
are bounded and holomorphic on $D_{R'}(0)$. Moreover, they coincide on $\R\cap D_{R'}(0) = (-R',R')$. Hence, they also coincide on $D_{R'}(0)$.
\end{proof}

Finally, one obtains the next proposition, and one refers to Fig. \ref{figd} for a graphic illustration.

\begin{prop}\label{pdr2}
Let $R > 0$ and $\widehat V \in \sinf ( {\rm L}^2({\mathbb T}) ) \cap {\mathcal A}_R (\widehat A_0)$, and let $R' > 0$ such that $2R ' = \min(R,\frac\pi8)$. Then, for any $\theta\in D_{R'}(0)$, we have
\begin{enumerate}
\item[(a)] $\sigma ( \widehat \la(\theta) )$ depends only on $\im (\theta)$.
\item[(b)] It holds: $\sigma_\mathrm{ess} ( \widehat \la(\theta) ) = \sigma_\mathrm{ess} ( \widehat \lao(\theta) ) = \sigma ( \widehat \lao(\theta) )$ and 
\begin{equation*}
\sigma ( \widehat \la(\theta) ) = \sigma_\mathrm{disc} ( \widehat \la(\theta) ) \bigsqcup \sigma_\mathrm{ess} ( \widehat \lao(\theta) ),
\end{equation*}
where the possible limit points of $\sigma_\mathrm{disc} ( \widehat \la(\theta) )$ belong to $\sigma_\mathrm{ess} ( \widehat \lao(\theta) )$.
\end{enumerate}
\end{prop}

\noindent \begin{proof}
Statement (a) is a consequence of the unitary equivalence established in Proposition \ref{pdr1}. Statement (b) follows from Lemma \ref{ld1}, the Weyl criterion on the invariance of the essential spectrum and \cite[Theorem 2.1, p. 373]{goh}.
\end{proof}

\subsubsection{Proof of  of Theorem \ref{t:tda}}

The proof of Statements (i) and (ii) follows now from a straightforward adaptation to our setting of the contents of \cite[Section 3.3]{bst} and \cite[Section 3.4]{bst} respectively.

\subsection{Resonances}\label{p:sr}

In this section the perturbation $V$ is assumed to satisfy Assumption \ref{ass:g+1}. The notations are those introduced in Section \ref{secd}. We will adopt the following principal determination of the complex square root: $\sqrt{\cdot} : {\mathbb C} \setminus (-\infty,0] \longrightarrow \big\lbrace z \in {\mathbb C} : \im(z) \ge 0 \big \rbrace$, and we set 
${\mathbb C}^+ := \big\lbrace z \in {\mathbb C} : \im(z) > 0 \big \rbrace$. 

\subsubsection{Definition of the resonances}

One defines the resonances of the operator $H_V$ near the spectral thresholds $\{0,4\}$. Notice that there is a simple way allowing to reduce the analysis near the second 
threshold $4$ to that of the first one $0$ (see \eqref{eq:s5,28}). 
Preliminary results will be firstly established. One recalls that $W_{\pm \gamma}$ denote the multiplication operators on $\ell^2(\bz)$ by the functions
$n \longmapsto {\rm e}^{\pm\frac{\gamma}{2}\vert n \vert}$.
One has the following lemma:

\begin{lem}\label{l:sdecU}
There exists  $\mathscr W \in \mathcal{B} \big( \ell^2(\bz,\mathscr{H}) \big)$ such that
$W = (W_{-\gamma} \otimes I) {\mathscr W} (W_{-\gamma} \otimes I)$.
In particular, one has
$
V = (\Gamma_1 W_{-\gamma} \otimes \Lambda_1) {\mathscr W} (W_{-\gamma} \Gamma_2 \otimes \Lambda_2).
$
Moreover, $V$ is compact.
\end{lem}

\noindent
\begin{proof}
Let ${\bf W}_{-\gamma}$ denote the multiplication operator by the function $n \longmapsto {\rm e}^{-\frac{\gamma}{2}\vert n \vert}$ on $\ell^2(\bz,\mathscr H)$.
Thus ${\bf W}_{-\gamma} = W_{-\gamma} \otimes I$ and one has
$$
W = (W_{-\gamma} \otimes I)  \mathscr W (W_{-\gamma} \otimes I) \quad {\rm with} \quad
\mathscr W := \big( {\rm e}^{\frac{\gamma}{2}\vert n \vert} w(n,m) {\rm e}^{\frac{\gamma}{2}\vert m \vert} \big)_{(n,m) \in \bz^2},
$$
i.e. $(\mathscr W \phi)(n) = \sum_{m \in \bz} {\rm e}^{\frac{\gamma}{2}\vert n \vert} w(n,m) {\rm e}^{\frac{\gamma}{2}\vert m \vert} \phi(m)$ for any 
$\phi \in \ell^2(\bz,\mathscr H)$. Therefore, to get the first claim of the lemma, it suffices to show that $\mathscr W$ is a bounded operator as follows. Similarly to 
\eqref{eq:g+11}, $\mathscr W$ can be represented canonically by
$
\mathscr W = \sum_{n,m} | \delta_n \rangle \langle \delta_m | \otimes {\rm e}^{\frac{\gamma}{2}\vert n \vert} w(n,m) {\rm e}^{\frac{\gamma}{2}\vert m \vert}.
$
Then, one obtains from Assumption \ref{ass:g+1} (ii) that
$$
\| \mathscr W \| \le \sum_{n,m} \| | \delta_n \rangle \langle \delta_m | \|  \| {\rm e}^{\frac{\gamma}{2}\vert n \vert} w(n,m) {\rm e}^{\frac{\gamma}{2}\vert m \vert} \|_{\mathscr H}
\lesssim \sum_{n,m} {\rm e}^{-\frac{\gamma}{2}\vert n \vert} {\rm e}^{-\frac{\gamma}{2}\vert m \vert} < \infty,
$$
and the claim follows. The compactness of $V$ follows for instance by that of $W_{-\gamma} \Gamma_2 \otimes \Lambda_2$, since the operator 
$W_{-\gamma} \otimes \Lambda_2$ is compact.
\end{proof}

\medskip

Let ${\bf J}$ be the selfadjoint unitary operator defined by \eqref{eq:s5,26} and introduce the operator
\begin{equation}\label{eq:spr}
V_{\bf J} := {\bf J} V {\bf J}^{-1}.
\end{equation}
Under  (i) of Assumption \ref{ass:g+1}, $\bf J$ and ${\bf J}^{-1}$ commute with the operators $\Gamma_j W_{-\gamma} \otimes \Lambda_j$, $j = 1$, $2$. Then, an 
immediate consequence of Lemma \ref{l:sdecU} is the following corollary:

\begin{cor}\label{c:sdecU}
There exists $\mathscr W_{\bf J} \in \mathcal{B} \big( \ell^2(\bz,\mathscr{H}) \big)$ such that
$
V_{\bf J} = (\Gamma_1 W_{-\gamma} \otimes \Lambda_1) \mathscr W_{\bf J} (W_{-\gamma} \Gamma_2 \otimes \Lambda_2).
$
\end{cor}

In the sequel, when no confusion arises, one sets
\begin{equation}\label{eq:spr}
{\bf V} := V \: {\rm or} \: - V_{\bf J}.
\end{equation}


For further use, one recalls the following result established in \cite{bst}:

\begin{lem}{\cite[Lemma 4.1]{bst}}
\label{l-3,1}
Set $z(\lambda) := \lambda^2$. Then, there exists $0 < \varepsilon_0 \le \frac{\delta}{8}$ small enough such that the operator-valued function
$\lambda \mapsto W_{-\gamma} \big( L_0 - z(\lambda) \big)^{-1} W_{-\gamma}$
admits a holomorphic extension from $D_{\varepsilon_0}^\ast(0) \cap {\mathbb C}^+$ to $D_{\varepsilon_0}^\ast(0)$, with values in $\sinf \big( \ell^2({\mathbb Z}) \big)$.
\end{lem}

The next result, whose proof follows easily using Lemma \ref{l-3,1}, is crucial for our analysis.

\begin{lem}\label{ls3,1}
There exists $0 < \varepsilon_0 \le \frac{\gamma}{8}$ small enough such that the operator-valued function
$
\lambda \mapsto (W_{-\gamma} \otimes \Lambda_2) \big( H_0 - z(\lambda) \big)^{-1} (W_{-\gamma} \otimes I)
$
admits a holomorphic extension from $D_{\varepsilon_0}^\ast(0) \cap {\mathbb C}^+$ to $D_{\varepsilon_0}^\ast(0)$, with values in 
$\sinf \big( \ell^2({\mathbb Z},\mathscr H) \big)$.
\end{lem}

\noindent
\begin{proof}
According to \eqref{eq:g+3}, one has
$
(W_{-\gamma} \otimes \Lambda_2) \big(H_0 - z(\lambda)\big)^{-1} (W_{-\gamma} \otimes I)= W_{-\gamma} \big( L_0 - z(\lambda) \big)^{-1} W_{-\gamma} \otimes \Lambda_2.
$
Since the operator $\Lambda_2$ is compact, then the claim follows by Lemma \ref{l-3,1}.
\end{proof}

Now, it follows from the identity
$(H_{\bf V} - z)^{-1} \left( I + {\bf V} (H_0 - z)^{-1} \right) = (H_0 - z)^{-1}$
that
\begin{equation*}\label{eq:s5,18}
\begin{split}
& (W_{-\gamma} \otimes \Lambda_2) (H_{\bf V} - z)^{-1} (W_{-\gamma} \otimes I) \\
& = (W_{-\gamma} \otimes \Lambda_2) (H_0 - z)^{-1} (W_{-\gamma} \otimes I) \big( I + (W_{\gamma} \otimes I) {\bf V}(H_0 - z)^{-1} (W_{-\gamma} \otimes I) \big)^{-1}.
\end{split}
\end{equation*}
By combining Lemma \ref{l:sdecU}, Corollary \ref{c:sdecU}, Lemma \ref{ls3,1} and (i) of Assumption \ref{ass:g+1}, one obtains that the operator-valued 
function
$\lambda \longmapsto (W_{\gamma} \otimes I) {\bf V} \big( H_0 - z(\lambda) \big)^{-1} (W_{-\gamma} \otimes I)$
is holomorphic in $D_{\varepsilon_0}^\ast(0)$ with values in $\sinf \big( \ell^2(\bz,\mathscr H)  \big)$. Therefore, by the analytic Fredholm extension theorem, 
the operator-valued function
\begin{equation*}\label{eq5,21}
\lambda \longmapsto  \big( I + (W_{\gamma} \otimes I) {\bf V} \big( H_0 - z(\lambda) \big)^{-1} (W_{-\gamma} \otimes I) \big)^{-1}
\end{equation*}
admits meromorphic extension from $D_{\varepsilon_0}^\ast(0) \cap \bc^+$ to $D_{\varepsilon_0}^\ast(0)$. Introduce the Banach weighted spaces 
\begin{equation}\label{eq:s5,22}
\ell_{\pm \gamma}^{2}(\bz,\mathscr H) := \ell_{\pm \gamma}^{2}(\bz) \otimes \mathscr H.
\end{equation}
The following proposition follows:
\begin{prop}\label{p:s5,1} 
The operator-valued function
\begin{equation*}\label{eq:s5,23}
\lambda \longmapsto \big( H_{\bf V} - z(\lambda) \big)^{-1} \in  \mathcal{B} \left( \ell_{\gamma}^{2}(\bz,\mathscr H),\ell_{-\gamma}^{2}(\bz,\mathscr H) \right),
\end{equation*} 
admits a meromorphic extension from $D_{\varepsilon_0}^\ast(0) \cap \bc^+$ to $D_{\varepsilon_0}^\ast(0)$. This extension will be denoted 
$R_{\bf V} \big( z(\lambda) \big)$.
\end{prop}

It follows from Lemma \ref{ls3,1} and Assumption \ref{ass:g+1} (i) the following result:

\begin{lem}\label{l:s5,2} 
The operator-valued functions
\begin{itemize}
\item $\lambda \longmapsto {\bf T}_V \big( z(\lambda) \big) := {\mathscr W} (W_{-\gamma} \Gamma_2 \otimes \Lambda_2) \big( H_0 - z(\lambda) \big)^{-1} 
(\Gamma_1 W_{-\gamma} \otimes \Lambda_1)$,
\item $\lambda \longmapsto {\bf T}_{-V_{\bf J}} \big( z(\lambda) \big) := -{\mathscr W}_{\bf J} (W_{-\gamma} \Gamma_2 \otimes \Lambda_2) 
\big( H_0 - z(\lambda) \big)^{-1} (\Gamma_1 W_{-\gamma} \otimes \Lambda_1)$,
\end{itemize} 
admit holomorphic extensions from $D_{\varepsilon_0}^\ast(0) \cap \bc^+$ to $D_{\varepsilon_0}^\ast(0)$, with values in $\sinf \big( \ell^2(\bz,\mathscr H) \big)$.
\end{lem}

One can now define the resonances of the operator $H_V$ near the spectral thresholds $\{0,4\}$. Notice that in the next definitions, 
the quantity $Ind_{\mathcal C}(\cdot)$ is defined in the appendix by \eqref{eqa,2}.

\begin{de}\label{d:s5,1} 
We define the resonances of the operator $H_V$ near $0$ as the poles of the meromorphic extension $R_V(z)$ of the resolvent 
$(H_V - z)^{-1}$ in $\mathcal{B} \left( \ell_{\gamma}^{2}(\bz,\mathscr H),\ell_{-\gamma}^{2}(\bz,\mathscr H) \right)$. The multiplicity of a resonance 
$z_0 := z_0(\lambda) = \lambda^2$ is defined by 
\begin{equation}\label{eq:s5,25}
\textup{mult}(z_0) := Ind_{\mathcal C} \big( I +  {\bf T}_V \big( z_0(\cdot) \big) \big),
\end{equation}
where $\mathcal C$ is a small contour positively oriented containing $\lambda$ as the only point satisfying $z_0(\lambda)$ is a resonance of 
$H_V$.
\end{de}

As said above, to define the resonances of the operator $H_V$ near $4$, there is a reduction exploiting a simple relation between the two thresholds 
$\{0,4\}$. Indeed, since $J L_0 J^{-1} = -L_0 + 4$, then one has ${\bf J} H_0 {\bf J}^{-1} = -H_0 + 4$, which implies that
${\bf J} (H_V - z) {\bf J}^{-1} = -H_0 + V_{\bf J} + 4 - z$, so that
\begin{equation}\label{eq:s5,28}
\begin{split}
 {\bf J} (W_{-\gamma} & \otimes \Lambda_2) (H_V - z)^{-1} (W_{-\gamma} \otimes I) {\bf J}^{-1}  \\
& = - (W_{-\gamma} \otimes \Lambda_2) \big( H_{-V_{\bf J}} - (4 - z) \big)^{-1} (W_{-\gamma} \otimes I).
\end{split}
\end{equation}
Let us set $u := 4 - z$. 
Since $u$ is near $0$ for $z$ near $4$, then by using identity \eqref{eq:s5,28}, one can define the resonances of the operator $H_V$ near $4$ as the points 
$z = 4 - u$ with $u$ pole of the meromorphic extension of the resolvent
$- \big( H_{-V_{\bf J}} - u \big)^{-1} : \ell_{\gamma}^{2}(\bz,\mathscr H) \rightarrow \ell_{-\gamma}^{2}(\bz,\mathscr H)$
near $0$ similarly to Definition \ref{d:s5,1}. More precisely, one has:

\begin{de}\label{d:s5,2} 
We define the resonances of the operator $H_V$ near $4$ as the points $z = 4 - u$ with $u$ pole of the meromorphic extension $R_{-V_{\bf J}}(u)$
of the resolvent $\big( H_{-V_{\bf J}} - u \big)^{-1}$ in $\mathcal{B} \left( \ell_{\gamma}^{2}(\bz,\mathscr H),\ell_{-\gamma}^{2}(\bz,\mathscr H) \right)$ 
near $0$. The multiplicity of a resonance $z_4 := z_4(\lambda) = 4 - \lambda^2$ is defined by 
\begin{equation}\label{eq:s5,31}
\textup{mult}(z_4) := Ind_{\mathcal C} \, \big( I + {\bf T}_{-V_{\bf J}} \big( 4 - z_4(\cdot) \big) \big),
\end{equation}
where $\mathcal C$ is a small contour positively oriented containing $\lambda$ as the only point satisfying $4 - z_4(\lambda)$ is a pole of 
$R_{-V_{\bf J}}(u)$.
\end{de}

\begin{rem} 
The resonances $z_\mu(\lambda)$ near the spectral thresholds $\mu \in \{0,4\}$ are defined respectively in some two-sheets Riemann surfaces 
$\mathcal{M}_\mu$. The discrete eigenvalues of the operator $H_V$ near $\mu$ are resonances. Furthermore, the algebraic multiplicity \eqref{eq1,9} 
of a discrete eigenvalue coincides with its multiplicity as a resonance near $\mu$ respectively defined by \eqref{eq:s5,25} and \eqref{eq:s5,31}. This can be 
shown for instance as in \cite{bst}.
\end{rem} 

\subsubsection{Proof of Theorem \ref{t1s}}

The first step is to characterize the resonances near $0$ and $4$ in terms of characteristic values.

\begin{prop}\label{p:s3,2}
For $\lambda_1 \in D_{\varepsilon_0}^\ast(0)$, the following assertions are 
equivalent:
\begin{itemize}
\item[(i)] $z_0(\lambda_1) = \lambda_1^{2} \in \mathcal{M}_0$ is a resonance of $H_V$, 
\item[(ii)] $z_{0,1} = z_0(\lambda_1)$ is a pole of $R_V(z)$, 
\item[(iii)] $-1$ is an eigenvalue of ${\bf T}_{V} \big( z_0(\lambda_1) \big)$,
\item[(iv)] $\lambda_1$ is a characteristic value of  $I + {\bf T}_{V} \big( z_0(\cdot) \big)$. 
Moreover, thanks to \eqref{eq:s5,25}, the multiplicity of the resonance $z_0(\lambda_1)$ coincides with that of the characteristic
value $\lambda_1$.
\end{itemize}
\end{prop}

\noindent
\begin{proof}
$(i) \Longleftrightarrow (ii)$ is just Definition \ref{d:s5,1}, $(ii) \Longleftrightarrow (iii)$ is a consequence of the identity
\begin{equation*}
\begin{split}
\big( I + {\mathscr W} (W_{-\gamma} & \Gamma_2 \otimes \Lambda_2) (H_0 - z)^{-1} (\Gamma_1 W_{-\gamma} \otimes \Lambda_1) \big) \\
& \big( I - {\mathscr W} (W_{-\gamma} \Gamma_2 \otimes \Lambda_2) (H_V- z)^{-1} (\Gamma_1 W_{-\gamma} \otimes \Lambda_1) \big) = I
\end{split}
\end{equation*}
which follows by the resolvent equation, and $(iii) \Longleftrightarrow (iv)$ follows by Definition \ref{d,a1} and \eqref{eqa,2}.
\end{proof}

Similarly, one has following proposition: 

\begin{prop}\label{p:s3,3}
For $\lambda_0 \in D_{\varepsilon_0}^\ast(0)$, the following assertions are equivalent:
\begin{itemize}
\item[(i)] $z_4(\lambda_0) = 4 - \lambda_0^{2} \in \mathcal{M}_4$ is a resonance of $H_V$, 
\item[(ii)] $\tilde{z}_{4,0} = \tilde{z}_4(\lambda_0) = 4 - z_4(\lambda_0) = \lambda_0^2$ is a pole of $R_{-V_J}(u)$, 
\item[(iii)] $-1$ is an eigenvalue of ${\bf T}_{-V_J}(\tilde{z}_4(\lambda_0))$,
\item[(iv)] $\lambda_0$ is a characteristic value of $I + {\bf T}_{-V_J}(\tilde{z}_4 \big( \cdot) \big)$.
Moreover, thanks to \eqref{eq:s5,31}, the multiplicity of the resonance $z_4(\lambda_0)$ coincides with that of the characteristic
value $\lambda_0$.
\end{itemize}
\end{prop}

\noindent
\begin{proof}
It follows as above.
\end{proof}

The second step of the proof is to split the sandwiched resolvent ${\bf T}_{\bf V} \big( z(\lambda) \big)$, $z(\lambda) = \lambda^2$, into a sum of a singular 
part at $\lambda = 0$ and a holomorphic part in the open disk $D_{\varepsilon_0}(0)$. By \eqref{eq:g+3} and (i) of Assumption \ref{ass:g+1}, one has
\begin{equation}\label{eq:ps1}
(W_{-\gamma} \Gamma_2 \otimes \Lambda_2) \big( H_0 - z(\lambda) \big)^{-1} (\Gamma_1 W_{-\gamma} \otimes \Lambda_1) 
= \Gamma_2 W_{-\gamma} \big( L_0 - z(\lambda) \big) W_{-\gamma} \Gamma_1 \otimes \Lambda_2 \Lambda_1.
\end{equation}
From \cite{bst}, we know that the summation kernel of the operator $W_{-\gamma} \big( L_0 - z(\lambda) \big)^{-1} W_{-\gamma}$ is given by
$
{\rm e}^{-\frac{\gamma}{2} \vert n \vert} R_0 \big( z(\lambda),n - m \big) {\rm e}^{-\frac{\gamma}{2} \vert m \vert},
$
with
\begin{equation}\label{eq:s3,17}
\begin{split}
R_0 \big( z(\lambda),n - m \big) & = \frac{i{\rm e}^{i \vert n - m \vert 2\arcsin \frac{\lambda}{2}}}{\lambda \sqrt{4 - \lambda^2}} = \frac{i}{\lambda \sqrt{4 - \lambda^2}} +
\frac{i \left( {\rm e}^{i \vert n - m \vert 2\arcsin \frac{\lambda}{2}} - 1 \right)}{\lambda \sqrt{4 - \lambda^2}} \\
& = \frac{i}{2\lambda} + \alpha(\lambda) + \beta(\lambda),
\end{split}
\end{equation}
where the functions $\alpha$ and $\beta$ are defined by
\begin{equation*}
\alpha(\lambda) := i \left( \frac{1}{\lambda \sqrt{4 - \lambda^2}} - \frac{1}{2\lambda} \right) \quad {\rm and} \quad \beta(\lambda) := \frac{i \left( e^{i \vert n - m \vert 
2\arcsin \frac{\lambda}{2}} - 1 \right)} {\lambda \sqrt{4 - \lambda^2}}.
\end{equation*}
Note that $\alpha$ and $\beta$ can be extended to holomorphic functions in $D_{\varepsilon_0}(0)$. By \eqref{eq:s3,17}, for 
$\lambda \in D_{\varepsilon_0}^\ast(0)$ 
\begin{equation}\label{eq:s3,19}
\left( W_{-\gamma} \big( L_0- z(\lambda) \big)^{-1} W_{-\gamma} x \right) (n) = \sum_{m \, \in \, {\mathbb Z}} \frac{i{\rm e}^{-\frac{\gamma}{2} \vert n \vert} 
{\rm e}^{-\frac{\gamma}{2} \vert m \vert} x(m)}{2\lambda} + \big( {\bf A} (\lambda) x \big)(n),
\end{equation}
where the operator ${\bf A}(\lambda)$ is defined by  
\begin{equation*}\label{eq:s3,20}
\big( {\bf A} (\lambda) x \big)(n) := \sum_{m \, \in \, {\mathbb Z}} {\rm e}^{-\frac{\gamma}{2} \vert n \vert} \alpha(\lambda) {\rm e}^{-\frac{\gamma}{2} \vert m \vert}  
x(m) + \sum_{m \, \in \, {\mathbb Z}} {\rm e}^{-\frac{\gamma}{2} \vert n \vert} \beta(\lambda) {\rm e}^{-\frac{\gamma}{2} \vert m \vert}  x(m).
\end{equation*}
Introduce the rank-one operator $\Xi : \ell^2({\mathbb Z}) \longrightarrow {\mathbb C}$ by $\Xi := \langle {\rm e}^{-\frac{\gamma}{2} \vert \cdot \vert} |$ 
so that its adjoint $\Xi^\ast : {\mathbb C} \longrightarrow \ell^2({\mathbb Z})$ be given by $\big( \Xi^\ast(\eta) \big)(n) := \eta {\rm e}^{-\frac{\gamma}{2} \vert n \vert} $. 
By putting this together with \eqref{eq:s3,19} one gets
\begin{equation}\label{eq:s3,200}
\left( W_{-\gamma} \big( L_0 - z(\lambda) \big)^{-1} W_{-\gamma} x \right) (n) =  \frac{i \big( \Xi^\ast \Xi x \big)(n)}{2\lambda} + \big( {\bf A} (\lambda) x \big)(n).
\end{equation}
For simplification, let us set
$$
\widetilde{\mathscr W} := \begin{cases} 
{\mathscr  W} & \text{if } \quad {\bf V} = V, \\ 
-{\mathscr W}_{\bf J}  & \text{if  } \quad {\bf V} = - V_{\bf J}. 
\end{cases}
$$
Then, we deduce from \eqref{eq:s3,200} and the above computations the following result:

\begin{prop}\label{p:s4,1} 
Let $\lambda \in D_{\varepsilon_0}^\ast(0)$. Then, setting ${\bf M} := \frac{1}{\sqrt{2}} \Xi$, one has 
\begin{equation*}
{\bf T}_{\bf V} \big( z(\lambda) \big) = \frac{i \widetilde{\mathscr W}}{\lambda} \Gamma_2 {\bf M}^\ast {\bf M} \Gamma_1 \otimes \Lambda_2 \Lambda_1
+ \widetilde{\mathscr W} \Gamma_2 {\bf A}(\lambda) \Gamma_1 \otimes \Lambda_2 \Lambda_1.
\end{equation*}
Furthermore, $\lambda \mapsto \widetilde{\mathscr W} \Gamma_2 {\bf A}(\lambda) \Gamma_1 \otimes \Lambda_2 \Lambda_1$ is holomorphic in the open disk
$D_{\varepsilon_0}(0)$ with values in $\sinf \big( \ell^2({\mathbb Z},\mathscr H) \big)$. 
\end{prop}


\medskip

The third and last step of the proof is to apply Proposition \ref{p,a1} as follows:

\medskip

Let $\mu \in \lbrace 0,4 \rbrace$. Then, from Propositions \ref{p:s3,2} and \ref{p:s3,3} together with Proposition \ref{p:s4,1}, it follows that $z_\mu(\lambda)$ is a 
resonance of $H_V$ near $\mu$ if and only if $\lambda$ is a characteristic value of the operator
\begin{equation*}
I + {\bf T}_{\bf V} \big( z(\lambda) \big) = I + \frac{i \widetilde{\mathscr W}}{\lambda} \Gamma_2 {\bf M}^\ast {\bf M} \Gamma_1 \otimes \Lambda_2 \Lambda_1
+ \widetilde{\mathscr W} \Gamma_2 {\bf A}(\lambda) \Gamma_1 \otimes \Lambda_2 \Lambda_1.
\end{equation*}
Since $\widetilde{\mathscr W} \Gamma_2 {\bf A}(\lambda) \Gamma_1 \otimes \Lambda_2 \Lambda_1$ is holomorphic in $D_{\varepsilon_0}(0)$ with values in 
$\sinf \big( \ell^2({\mathbb Z},\mathscr H) \big)$ while $ i \widetilde{\mathscr W} \Gamma_2 {\bf M}^\ast {\bf M} \Gamma_1 \otimes \Lambda_2 \Lambda_1$ is 
finite-rank, then Theorem \ref{t1s} follows by applying Proposition \ref{p,a1} with $\mathcal{D} = D_r(0)$, $Z = \lbrace 0 \rbrace$, and 
$F = I + {\bf T}_{\bf V} \big( z(\cdot) \big)$.

\section{Appendix}\label{sa}

\subsection{The ${\mathcal A}({\bf A}_0)$ class}\label{se}

One refers to \cite[Chapter III]{kato} for general considerations on bounded operator-valued analytic maps. The next lemma provides examples of analytic vectors for ${\bf A}_0$.

\begin{lem}\label{l:se1}
For any $n \in \bz$ and any vector $u \in \mathscr{H}$, $\delta_n \otimes u$ is an analytic vector for ${\bf A}_0$.
\end{lem}
\noindent \begin{proof}
For any $\theta \in D_\frac{1}{2}(0)$, it follows from \cite[Lemma 6.2]{bst} that the power series
$$
\sum_{k = 0}^\infty \frac{\vert \theta \vert^k}{k!} \big\Vert {\bf A}_0^k (\delta_n \otimes u) \big\Vert
= \sum_{k = 0}^\infty \frac{\vert \theta \vert^k}{k!} \big\Vert (A_0^k \otimes I) \delta_n \otimes u \big\Vert 
= \sum_{k = 0}^\infty \frac{\vert \theta \vert^k}{k!} \big\Vert A_0^k \delta_n \big\Vert \Vert u \Vert 
$$
converges. This concludes the proof.
\end{proof}

In the continuity of (i) of Remark \ref{r:sae}, one has the following result:
\begin{prop}\label{p:se1}
Let $(n,m) \in \bz^2$ and $W_{nm}$ be the operator defined by \eqref{eq:g+10}. Then, one has $W_{nm} \in {\mathcal A}_{\frac{1}{2}}({\bf A}_0)$.
The holomorphic extension map is given by $D_\frac{1}{2}(0) \ni \theta \longmapsto \vert {\rm e}^{i\theta A_0} \delta_n \rangle \langle {\rm e}^{i{\bar \theta} A_0} \delta_m \vert \otimes w(n,m)$. In particular, any finite linear combination of such $W_{nm}$ belongs to ${\mathcal A}_{\frac{1}{2}}({\bf A}_0)$.
\end{prop}
\noindent \begin{proof}
This is an immediate consequence of \cite[Lemma 6.2]{bst}.
\end{proof}

Now, we aim at proving that if the perturbation $V$ satisfies Assumption \ref{ass:g+1}, with $\Gamma_j$ proportional to $I$ for $j = 1,2$, then it belongs to ${\mathcal A}_{R_{\gamma}}({\bf A}_0)$ for some $R_\gamma > 0$ (see Proposition \ref{pe3}).

Referring to \cite[Section 3.1]{bst} for the details, we recall that for any $\psi \in \mathrm{L}^2(\mathbb{T}),$
\begin{equation}\label{eq6,4}
( \e^{i\theta {\widehat A}_0} \psi ) (\alpha) = \psi ( \varphi_\theta(\alpha)) \sqrt{J(\varphi_\theta)(\alpha)}\, , \alpha \in {\mathbb T}
\end{equation}
where
\begin{itemize}
\item $(\varphi_\theta)_{\theta\in\R}$ is the flow solution of the system:
$
\begin{cases} 
\partial_{\theta} \varphi_\theta (\alpha) = 2\sin \big( \varphi_\theta(\alpha) \big), \\ 
\varphi_0(\alpha) = \mathrm{id}_\mathbb{T}(\alpha) = \alpha \: \: \textup{for each} \: \: \alpha \in \mathbb{T},
\end{cases}
$
\item $J(\varphi_\theta)(\alpha)$ denotes the Jacobian of the transformation $\alpha \mapsto \varphi_\theta (\alpha)$.
\end{itemize}
Explicitly, one has:
$$
\varphi_\theta(\alpha) = \pm \arccos \left( \frac{-{\rm th}(2\theta) + \cos \alpha}{1 - {\rm th}(2\theta) \cos \alpha} \right)
$$
for $\pm \alpha \in \mathbb{T}$. By using \eqref{eq6,4} and the fact that $\varphi_{\theta_1} \circ \varphi_{\theta_2} = \varphi_{\theta_1 + \theta_2}$ for all $(\theta_1, \theta_2) \in \R^2$, one gets 
for all $\theta \in \R$
\begin{equation*}
( \e^{i\theta {\widehat A}_0} \widehat L_0 \e^{-i\theta {\widehat A}_0}\psi )(\alpha) = f ( \varphi_\theta(\alpha)) \psi(\alpha) .
\end{equation*}
Paraphrasing \cite[Lemma 6.3]{bst}, we have that:
\begin{lem}\label{le3}
There exists $0 < R_0 < 1/2$ such that for any $0<R<R_0,$
\begin{equation}
\sup_{n\in {\mathbb Z}} \left(\sup_{\theta \in \overline{D_R(0)}} \big\Vert {\rm e}^{i\theta A_0} \delta_n \big\Vert 
{\rm e}^{- \vert n \vert \sup_{(\theta,\alpha) \in \overline{D_R(0)} \times {\mathbb T}} \vert \im(\varphi_\theta(\alpha)) \vert} \right)< \infty .
\end{equation}
\end{lem}

\begin{prop}\label{pe3}
Let $V$ satisfy Assumption \ref{ass:g+1} with $\Gamma_j = \mu_j I$, $j = 1$, $2$, $\mu_j \in \bc$. Then, there exists $R_\gamma > 0$ such that $V$ 
belongs to ${\mathcal A}_{R_\gamma}({\bf A}_0)$. The extension map being given by
$$
D_{R_\gamma}(0) \ni \theta \longmapsto {\rm e}^{i\theta {\bf A}_0} V {\rm e}^{-i\theta {\bf A}_0} = 
\mu_1 \mu_2 \,\sum_{n,m}  \vert {\rm e}^{i\theta A_0} \delta_n \rangle \langle {\rm e}^{i{\bar \theta} A_0} \delta_m \vert \otimes \Lambda_1 w(n,m) \Lambda_2.
$$
\end{prop}
\noindent \begin{proof}
Thanks to Remark \ref{r:spd} (ii), one has 
$$
V = \mu_1 \mu_2 \,\sum_{n,m} (I \otimes \Lambda_1) W_{n,m} (I \otimes \Lambda_2)
= \mu_1 \mu_2 \,\sum_{n,m} \vert \delta_n \rangle \langle \delta_m \vert \otimes \Lambda_1 w(n,m) \Lambda_2.
$$
Now as in \cite[Proposition 6.4]{bst}, with the help of Lemma \ref{le3} one can pick $0 < R_{\gamma} < \frac{1}{2}$ small enough such that
$\sup_{\theta \in \overline{D_{R_\gamma}(0)}} \big\Vert {\rm e}^{i\theta A_0} \delta_n \big\Vert \le C {\rm e}^{\frac{\delta}{2}\vert n \vert}$, for some constant $C > 0$
(independent of $n$). Therefore, it follows that
\begin{align*}
\sum_{n,m} \sup_{\theta \in \overline{D_{R_\gamma}(0)}} & \big\Vert {\rm e}^{i\theta {\bf A}_0} (I \otimes \Lambda_1) W_{n,m} (I \otimes \Lambda_2) {\rm e}^{-i\theta {\bf A}_0} \big\Vert \\
& = \sum_{n,m} \sup_{\theta \in \overline{D_{R_\gamma}(0)}} \big\Vert ({\rm e}^{i\theta A_0} \otimes I) (I \otimes \Lambda_1) (|\delta_n \ket \bra \delta_m | \otimes w(n,m)) (I \otimes \Lambda_2) ({\rm e}^{-i\theta A_0} \otimes I) \big\Vert \\
& = \sum_{n,m} \sup_{\theta \in \overline{D_{R_\gamma}(0)}} \big\Vert {\rm e}^{i\theta A_0} \delta_n \big\Vert
\big\Vert {\rm e}^{i{\bar \theta} A_0} \delta_m \big\Vert \Vert \Lambda_1 w(n,m) \Lambda_2 \Vert \\
& \lesssim \sum_{n,m} {\rm e}^{\frac{\gamma}{2} (\vert n \vert + \vert m \vert)} \Vert \Lambda_1 w(n,m) \Lambda_2 \Vert < \infty.
\end{align*}
By arguing as in \cite[Proposition 6.5]{bst}, one gets the claim.
\end{proof}

One concludes this section by the following remark:

\begin{rem} \label{rfin}
\begin{enumerate}
\item[{\rm (i)}] If $w(n,m) \in \sinf({\mathscr H})$ for some $(n,m) \in \bz^2$, then $W_{n,m} \in \sinf \big( \ell^2({\mathbb Z},\mathscr H) \big)$.
\item[{\rm (ii)}] If $w(n,m) \in \sinf({\mathscr H})$ for all $(n,m) \in \bz^2$ and if moreover $\sum_{n,m} \Vert w(n,m) \Vert < \infty$, then
$W \in \sinf \big( \ell^2({\mathbb Z},\mathscr H) \big)$. Indeed, for all $(n,m) \in \bz^2$, $\Vert W_{n,m} \Vert = \Vert \delta_n \Vert \Vert \delta_m \Vert \Vert w(n,m) \Vert = \Vert w(n,m) \Vert$, so that $W$ is the limit of an absolutely convergent series of compact operators.
\end{enumerate}
\end{rem}

\subsection{Characteristic values}\label{cv}

We recall some tools we need on characteristic values of finite meromorphic operator-valued functions. For more details on the subject, one refers 
to \cite{go} and the book \cite[Section 4]{goh}. The content of this section follows \cite[Section 4]{goh}. Let $\mathscr{H}$ be Hilbert space as above.

\begin{de}
Let $\mathcal{U}$ be a neighborhood of a fixed point $w \in \bc$, and 
$F : \mathcal{U} \setminus \lbrace w \rbrace \longrightarrow {\mathcal B}(\mathscr{H})$ 
be a holomorphic operator-valued function. The function $F$ is said to be finite 
meromorphic at $w$ if its Laurent expansion at $w$ has the form
$F(z) = \sum_{n = m}^{+\infty} (z - w)^n A_n$, $m > - \infty$,
where (if $m < 0$) the operators $A_m, \ldots, A_{-1}$ are of finite rank.
Moreover, if $A_0$ is a Fredholm operator, then the function $F$ is said to be Fredholm 
at $w$. In that case, the Fredholm index of $A_0$ is called the Fredholm index of $F$ 
at $w$.
\end{de}

One has the following proposition:

\begin{prop}{\cite[Proposition 4.1.4]{goh}}\label{p,a1}
Let $\mathcal{D} \subseteq \mathbb{C}$ be a connected open set, $Z \subseteq \mathcal{D}$ be a closed and discrete subset of $\mathcal{D}$, and 
$F : \mathcal{D} \longrightarrow {\mathcal B}(\mathscr{H})$ be a holomorphic operator-valued function in $\mathcal{D} \backslash Z$. Assume that:
$F$ is finite meromorphic on $\mathcal{D}$ (i.e. it is finite meromorphic near each point of $Z$), $F$ is Fredholm at each point of $\mathcal{D}$, 
there exists $w_0 \in \mathcal{D} \backslash Z$ such that $F(w_0)$ is invertible. 
Then, there exists a closed and discrete subset $Z'$ of $\mathcal{D}$ such that:
$Z \subseteq Z'$, $F(z)$ is invertible for each $z \in \mathcal{D} \backslash Z'$,
$F^{-1} : \mathcal{D} \backslash Z' \longrightarrow {\rm GL}(\mathscr{H})$ is finite
meromorphic and Fredholm at each point of $\mathcal{D}$.
\end{prop}

\noindent
In the setting of Proposition \ref{p,a1}, one defines the characteristic values of $F$ and their multiplicities:

\begin{de}\label{d,a1}
The points of $Z'$ where the function $F$ or $F^{-1}$ is not holomorphic are called the
characteristic values of $F$. The multiplicity of a characteristic value $w_0$ is 
defined by
\begin{equation}
{\rm mult}(w_0) := \frac{1}{2i\pi} \textup{Tr} \oint_{\vert w - w_0 \vert = \rho} 
F'(z)F(z)^{-1} dz,
\end{equation}
where $\rho > 0$ is chosen small enough so that $\big\lbrace w \in \bc : \vert w - 
w_0 \vert \leq \rho \big\rbrace \cap Z' = \lbrace w_0 \rbrace$.
\end{de}

\noindent
According to Definition \ref{d,a1}, if the function $F$ is holomorphic in $\mathcal{D}$,
then the characteristic values of $F$ are just the complex numbers $w$ where the operator 
$F(w)$ is not invertible. Then, results of \cite{go} and \cite[Section 4]{goh} imply that 
${\rm mult}(w)$ is an integer.
Let $\Omega \subseteq \mathcal{D}$ be a connected domain with boundary $\partial \Omega$ 
not intersecting $Z'$. The sum of the multiplicities of the characteristic values of
the function $F$ lying in $\Omega$ is called {\it the index of $F$ with respect to the 
contour $\partial \Omega$} and is defined by 
\begin{equation}\label{eqa,2}
Ind_{\partial \Omega} \hspace{0.5mm} F := \frac{1}{2i\pi} \textup{Tr} 
\oint_{\partial \Omega} F'(z)F(z)^{-1} dz = \frac{1}{2i\pi} \textup{Tr} 
\oint_{\partial \Omega} F(z)^{-1} F'(z) dz.
\end{equation}

\medskip

\noindent
{\bf Acknowledgements:} O. Bourget is supported by the Chilean Fondecyt Grant $1161732$. D. Sambou is supported by the Chilean Fondecyt Grant $3170411$. A. Taarabt is supported by the Chilean Fondecyt Grant $11190084$.

\end{document}